# A comparison of different methods of identifying publications related to the United Nations Sustainable Development Goals: Case Study of SDG 13 – Climate Action

Philip J. Purnell
[1]Centre for Science and Technology Studies,
Leiden University,
P.O. Box 905, 2300 AX Leiden, The Netherlands
[2]United Arab Emirates University, Al Ain, UAE
Tel: +971 50 552 9356
p.j.purnell@cwts.leidenuniv.nl
ORCID: 0000-0003-3146-2737

## Abstract

As sustainability becomes an increasing priority throughout global society, academic and research institutions are assessed on their contribution to relevant research publications. This study compares four methods of identifying research publications related to United Nations Sustainable Development Goal 13 – climate action. The four methods, Elsevier, STRINGS, SIRIS, and Dimensions have each developed search strings with the help of subject matter experts which are then enhanced through distinct methods to produce a final set of publications. Our analysis showed that the methods produced comparable quantities of publications but with little overlap between them. We visualised some difference in topic focus between the methods and drew links with the search strategies used. Differences between publications retrieved are likely to come from subjective interpretation of the goals, keyword selection, operationalising search strategies, AI enhancements, and selection of bibliographic database. Each of the elements warrants deeper investigation to understand their role in identifying SDG-related research. Before choosing any method to assess the research contribution to SDGs, end users of SDG data should carefully consider their interpretation of the goal and determine which of the available methods produces the closest dataset. Meanwhile data providers might customise their methods for varying interpretations of the SDGs.

## 1. Introduction

### 1.1. UN Sustainable Development Goals

The United Nations described a set of sustainable development goals (SDGs) within its 2030 sustainable development agenda. These goals were launched on 1$^{st}$ January 2016 and will be in place until 2030. The agenda includes 17 SDGs that are associated with 169 targets and progress is to be measured using 232 indicators (United Nations, 2017). The goals urge political, scientific, economic, and societal change to address global challenges and ensure sustainable development of the planet and all its inhabitants. To achieve these goals, all sectors of society are expected to participate including higher education institutions and research centres.

Table 1. United Nations SDG 13 goals and targets

| SDG | 13 |
|---|---|
| Short name | Climate action |
| Long name | Take urgent action to combat climate change and its impacts |
| Targets | 13.1 Strengthen resilience and adaptive capacity to climate-related hazards and natural disasters in all countries<br><br>13.2 Integrate climate change measures into national policies, strategies and planning<br><br>13.3 Improve education, awareness-raising and human and institutional capacity on climate change mitigation, adaptation, impact reduction and early warning<br><br>13.A Implement the commitment undertaken by developed-country parties to the United Nations Framework Convention on Climate Change to a goal of mobilizing jointly $100 billion annually by 2020 from all sources to address the needs of developing countries in the context of meaningful mitigation actions and transparency on implementation and fully operationalize the Green Climate Fund through its capitalization as soon as possible<br><br>13.B Promote mechanisms for raising capacity for effective climate change-related planning and management in least developed countries |

| | and small island developing States, including focusing on women, youth and local and marginalized communities |

Source: United Nations resolution A/RES/71/313

One key step in assessing progress of the academic community against the SDGs is to identify the relevant research outputs. These are usually articles published in scholarly journals and books, or presentations at conferences. Research publications are indexed in large databases which can be searched using strings of keywords. If the search terms match words in the article title or abstract, then that article is included in the search results. In this paper, we compare different methods of identifying research publications related to SDGs. Our focus is on SDG 13: Climate action, whose goals and targets are shown in Table 1. We chose SDG 13 because it affects the entire global population and environment and is strongly dependent on scholarly research.

As research into the SDGs develops, the number of efforts to create search strings grows, each different from the others. Current methods of which we are aware are summarised in Table 2.

Table 2. Current methods of defining SDG-related research

| Group | Data source | Method |
| --- | --- | --- |
| Elsevier 2020 | Scopus | Boolean search strings |
| Elsevier 2021 | Scopus | ML-enhanced |
| Bergen | Web of Science | Boolean search strings |
| Aurora (Elsevier 2019) | Scopus | Boolean search strings |

| Clarivate ISI | Web of Science | Citation enhanced |
| --- | --- | --- |
| STRINGS | Web of Science | Citation clustering-enhanced |
| SIRIS Academic | Various | ML-enhanced |
| Digital Science | Dimensions | ML-enhanced |

## 1.2. Study aims

This study aims to quantify the different datasets produced by following four methods and shed light on the underlying causes of those differences. Each of the methods selected for this study has made some attempt to enhance their datasets through algorithms. These were done in different ways and sometimes at different stages of the process. Although the machine learning element was not scrutinised in this paper, it is worth highlighting the differences because they are likely to influence the resulting datasets. The following methods were chosen for the study because they cover all SDGs, and we have access to the search terms used and resulting publications:

- Elsevier (2021) – used to calculate part of the 2021 Impact Rankings (Times Higher Education, 2020)
- STRINGS – Steering Research and Innovation for Global Goals (Confraria, Noyons, & Ciarli, 2021)
- SIRIS Academic – a European consulting firm (SIRIS Academic, 2020)
- Dimensions – developed by Digital Science (Wastl, Hook, Fane, Draux, & Porter, 2020)

It is important to point out that we did not attempt to evaluate the accuracy of the methods or to pick a winner. We deliberately chose not to develop our own method because of the multitude of questions raised when defining a ground truth (Gläser, Glänzel, & Scharnhorst, 2017). Our intention was to shed light on the discrepancies produced when applying different perspectives to the same question.

We did not use other published methods because (e.g.):

- Bergen: We could not run the complex search in our version of Scopus and there were too many records to export from the Scopus database
- Aurora: The method was not fully developed for global analysis
- Clarivate ISI: The publications are not assigned to individual SDGs

Specifically, we analyse the search strings (inputs) used by each method, and sub-classify them into general terms, policy-related terms, and technical terms. The use of subject matter experts will surely influence the type of search terms used and consequently determine the set of research publications identified as related to SDG 13.

We then compare the size of the resulting sets of publications identified by each of the four methods. We perform quantitative comparison of the overlap and surplus of each of the methods. We then discuss the influence of the type of keywords used in the search strings in determining the final dataset.

Finally, we compare the articles identified by the different methods (outputs) using VOSviewer maps. These maps help visualise the nuances of each of the methods and show the links with the corresponding search strategies.

Research questions:

1. To what extent do different search strategies produce different sets of SDG13-related publications?
2. What is the impact of including different types of search terms in the search strings?
3. What is the impact of using larger, more inclusive data sources over smaller, more selective ones?

### 1.3. Assessing university impact

This study is important because universities are increasingly asked to demonstrate their 'impact' on society in areas such as sustainability, so there is an increasing need to expand the definition of university performance to encompass the area of societal impact. University contribution towards the SDGs is therefore both welcomed and expected by their stakeholders and society in general. This expectation is accompanied by efforts to measure progress against the SDGs using performance indicators appropriate for universities.

Academic publications are frequently used in research evaluation and universities are routinely assessed on their article output for internal performance review and international benchmarking. Research articles related to the SDG goals and targets are therefore an appropriate unit upon which to base such assessments. Indeed, there is now a global ranking of universities based on their progress against the SDGs, about a quarter of which is based on their research publications related to the SDGs (Times Higher Education, 2021b). The first two editions of this ranking assessed universities on their Scopus-indexed publications retrieved via a series of search strings. The 2021 edition further extended the publication datasets through a process of machine learning.

Research publications are typically analysed using large international multidisciplinary databases comprised of scholarly research papers. Bibliographic databases such as Scopus,

Web of Science, and Dimensions include journal articles, conference proceedings, and research published in books. However, they each have their own selection and coverage policy that results in differences between the publications included. Therefore, the choice of bibliographic database will determine the resulting dataset depending on the selection and coverage policy of the database. Running the same search in different databases will yield different results.

## 1.4. Search strategies

Even using the same data source does not make SDG-related publication datasets comparable, indeed it has been shown (Armitage, Lorenz, & Mikki, 2020a) that differences in search strategies make a big difference in outcomes. For SDG 13: Climate action, only about one-third of articles were found by two different approaches.

In bibliographic searching, the search strategy is of key importance in determining the final set of publications. The search for SDG-related research is in its infancy and we aim to advance current understanding of the relationship between different search strategies and the resulting publication datasets. The UN described the goals, targets and indicators using specific terms, and the UN and other bodies have published related documents and reports also using subject-specific language. As these reports were written by people with close subject knowledge, they can be used as sources of search terms in a bibliographic database. Subject matter experts can then refine the searches to improve their recall and precision.

In terms of information retrieval, recall is the number of relevant publications retrieved as a share of all the relevant publications. To maximise recall, one would make the search strategy as broad as possible and not be concerned with the prospect of finding false positives among the results. Meanwhile, precision is the number of relevant publications retrieved as a share of all retrieved publications. To maximise precision, the search strategy should be as narrow as possible to exclude any irrelevant publications. The trick in identifying SDG-related research is for the search strings to be both effective at recall while remaining precise.

In order to assess recall, it is first necessary to define a precise, yet representative, set of reference publications. The way a method operationalises its interpretation of the SDGs, through a reference dataset used to measure recall, will largely influence the results and overlap with other methods.

## 2. Literature review

### 2.1. Current methods of defining SDG research

One of the earliest insights into sustainability science was developed by the publisher Elsevier in collaboration with SciDev.Net (Elsevier & SciDev.Net, 2015). This was the first in a series of reports that aimed to describe the research landscape in areas related to the SDGs. In order to identify research papers related to the SDGs, Elsevier worked with field experts to design sets of Boolean queries that were applied to the Scopus Advanced search (Jayabalasingham, Boverhof, Agnew & Klein, 2019). The keywords were related to research themes linked to the six Essential Elements (Dignity, People, Prosperity, Planet, Justice, and Partnership) described by the United Nations (2014). The 17 SDGs are grouped around these six Essential Elements. The experts identified key phrases from the titles and abstracts of relevant reports and those keywords were then used to search for scholarly articles indexed in the Scopus database. The advantage of this method was its ability to retrieve papers that use specific terms related to various aspects of one or more of the SDGs, without having to explicitly use the term 'sustainable development goal'.

A team of library and information specialists from the University of Bergen in Norway set out to discover the degree to which the design of the search strings affected the resulting set of publications (Armitage et al., 2020a). They developed a set of complex Boolean search strings (Armitage, Lorenz, & Mikki, 2020b) for selected SDGs and queried them in Web of Science Advanced Topic Search. They then translated the 2019 Elsevier Boolean queries (Jayabalasingham et al., 2019) into Web of Science search strings and compared the results with their own 'Bergen' set. Comparison showed only a quarter of records were returned by both the 2019 Elsevier and the Bergen search strings with large quantities of publications not retrieved when using one or the other set of terms. The authors concluded that even mildly modifying the search strings used for specific SDGs will significantly change the resulting set of academic papers found. The Bergen group also addressed the question about what constitutes SDG-related research by creating two sets of search terms for each SDG; those relevant to the topic of e.g., clean water and sanitation, known as the Bergen Topic Approach (BTA), and papers on efforts to actually combat the challenges described by the SDGs, known as the Bergen Action Approach (BAA). The topic approach retrieved larger datasets than the action related searches, although the degree of overlap of these two approaches with the Elsevier dataset varied by SDG.

The SDG targets can be vague, weak, or non-essential (International Science Council, 2015) which makes it unclear which words or phrases in a target should be used in a search. Even search strings using the same initial terms will produce different results depending on how they are refined. The Bergen group reported that their queries tended to use more combinations of terms requiring each to be included in a paper for it to be returned in the results. For example, the Bergen group required the term 'climate change' to be combined with other terms found in the SDG 13 targets such as 'adaptation' or 'mitigation', whereas the 2019 Elsevier search would return results due to the simple appearance of the term 'climate change'. On the other hand, the 2019 Elsevier strategy refined its final dataset by excluding any papers that contain the term 'drug' or 'geomorphology'. Such papers relate to medicine and changes in earth layers related to prehistoric climate changes rather than those related to modern day climate action (Jayabalasingham et al., 2019). Both these methods aim to refine the dataset but will obviously lead to differences in results.

Since the comparison with the Bergen method, the complexity of Elsevier's search strategy has increased considerably. The breadth of Boolean queries has expanded to capture a wider range of related concepts such as carbon capture/mitigation, $CO_2$ in combination with global warming, or environmental impacts. The sole appearance of the term 'climate change' is no longer sufficient to retrieve publications. Similarly, the exclusion criteria in the 2021 method have been refined to over 30 specific terms, replacing the two in the 2019 method. Elsevier has published a full description of their methods along with the search strategies (Rivest et al., 2021).

Science-Metrix, now part of Elsevier, has described how analysts who are familiar with the SDG targets have defined sets of seed keywords for each SDG (Provençal, Campbell, & Khayat, 2021; Rivest et al., 2021). In this scenario, the preference for precision over recall is emphasised meaning that the dataset is expected to contain publications with high relevance to the SDG targets, even at the expense of missing some.

Several of the methods including Bergen, SIRIS and Dimensions aimed to capture phrases used in context rather than only in their exact form by using proximity searches, so that 'climate impact'~3 would also capture phrases such as 'climate change impact' and 'changing climate and its impact on health'. Again, these methods used different levels of proximity and on different search terms so the effect would of course compound differences in the publications retrieved.

Other groups have also developed Boolean search queries to describe bodies of SDG-related research. For example Jetten, Veldhuizen, Siebert, Ommen Kloeke, & Darroch (2019) aimed to discover the extent to which Wageningen University's work to improve food security through innovative technologies influenced media and policy documents. Similarly, Körfgen et al. (2018) developed a detailed keyword catalogue which found nearly a fifth of Austrian universities' research output was related to the SDGs. An attempt to assess Spanish public universities' contribution to the SDGs (Blasco, Brusca, & Labrador, 2021) used a composite indicator that included the Times Higher Education Impact rankings which is in turn partially based on Elsevier's 2021 keyword search string.

The Aurora Network of universities created an initial classification model to enable them to identify which research publications were related to each SDG and whether these influenced government policy (Vanderfeesten & Otten, 2017). They began with a strict version that was limited to keywords found in UN policy documents that described the goals, targets, and indicators. A number of subsequent versions of these search terms gradually added more keywords including synonyms, new terms from updated UN documents, keyword combinations, and terms retrieved though survey data (Vanderfeesten, Spielberg, & Gunes, 2020), Elsevier (Jayabalasingham et al., 2019; Rivest et al., 2021), and SIRIS Academic (Duran-Silva, Fuster, Massucci, & Quinquillà, 2019). The Aurora bibliometric tool queries the Scopus database and has been used by the Association of Dutch Universities (VSNU) to create a sustainability impact dashboard (Association of Dutch Universities, 2019).

Clarivate has used a technique known as bibliographic coupling to approach the problem (Nakamura, Pendlebury, Schnell, & Szomszor, 2019). The Clarivate method identifies any paper in Web of Science that has used the term 'sustainable development goal' in the title, abstract, or keywords and defines them as 'core' papers. They then add to these any paper that has cited one or more core papers. The citing papers plus the core papers makes up the SDG-related dataset.

The Science Policy & Research Unit at Sussex University and the United Nations Development Programme are leading a collaboration of several research centres known as STRINGS – Steering Research and Innovation for Global Goals. This collaboration has taken a novel approach in order to discover whether SDG research priorities in certain countries match those in which the related socio-economic challenges are greatest. They selected seed terms from a broad range of policy, technical, and scientific reports along with web forums and official UN documents using a combination of algorithms and expert opinion (Confraria et al., 2021) in an attempt to capture terms used by a broad section of society. They also compared their search strings with those used by Bergen and SIRIS Academic to remove false negatives from their results (Rafols, Noyons, Confraria, & Ciarli, 2021). The resulting combinations of terms associated with each SDG were then searched in Web of Science and used to identify clusters of SDG-related publications. If a certain proportion of publications were retrieved from one cluster, then the whole cluster was added to the dataset. If the threshold was not reached, the cluster was not added. These clusters group publications that are related by citation links and this offers a way to identify not only SDG-related publications that use specific search terms but also SDG-related publications that do not use these search terms but that have citation links to publications that do use the search terms.

SIRIS Academic, a European consulting firm has looked through a broader set of document types including R&D projects hosted on the Community Research and Development Information Service (CORDIS) (SIRIS Academic, 2020). This repository comprises primary results from European Union funded Framework Programme projects ranging from FP1 to Horizon 2020 (European Commission, 2020).

Digital Science has developed an approach that queries the Dimensions database. The results were analysed by country and the proportion of national output calculated for each of the SDGs. A similar proportion in each SDG was considered a well-rounded footprint, while diverse emphasis was considered a skewed profile. Digital Science has also attempted to establish the extent of international collaboration for each SDG and to map the SDGs onto established scientific fields. The Dimensions SDG data have been used in a study by the Nature Index of leading science cities (Nature, 2021).

## 2.2. Algorithmic enhancements

An emerging trend is to employ machine learning to enrich datasets of SDG-related publications. In this model, manually selected keywords are used to identify a set of seed papers from a bibliographic data source. An AI algorithm then learns from these seed papers to recognise other relevant publications.

In 2021, the Elsevier team enriched its 2020 dataset through machine learning adding approximately 10% to its dataset by improving recall. They used the title, keywords, key descriptor terms, journal subject area, and abstract from around 1 million publications related to SDGs to create a computer algorithm that elicited records relevant to each of the SDGs through a machine learning model (Rivest et al., 2021). Times Higher Education used these results in the calculation of the 2021 Impact Rankings (Times Higher Education, 2020).

SIRIS has created a controlled vocabulary for each SDG defining its 'semantic breadth' through a manual process of reading reports and identifying seed keywords (SIRIS Academic, 2020). As a second step, they have used deep learning to train a neural network model to find synonyms with the seed keywords and create an ontology. The ontology is then matched with terms logically linked with the seed keywords in the CORDIS repository. A final quality check comprised human revision of results generated by the automated method for relevance to the original definition of the SDGs.

Digital Science's machine learning approach (Wastl et al., 2020) involved generating 17 training sets and using natural language processing to create an SDG classification scheme searchable in Dimensions (Wastl et al., 2020).

### 2.3. Bibliographic data sources

Since the 1970s, Web of Science and its components have been routinely used for evaluating journal impact (e.g., Garfield, 1972), university benchmarking (e.g., van Raan, 1999), national research impact assessment (Adams, 1998), the contribution of individual researchers (Hirsch, 2005) and the development of advanced bibliometric indicators (e.g., Waltman, van Eck, van Leeuwen, Visser, & van Raan, 2011).

Elsevier launched Scopus, its global abstract and citation database of research papers from scholarly books, scientific conferences, and academic journals in 2004. Scopus has gradually become a key data source used for bibliometric studies of research output (Archambault, Campbell, Gingras, & Lariviere, 2009; Baas, Schotten, Plume, Côté, & Karimi, 2020; Schotten, Aisati, Meester, Steiginga, & Ross, 2017). Recently, Digital Science's Dimensions has also become an interesting data source for bibliometric studies (Herzog, Hook, & Konkiel, 2020; Hook, Porter, & Herzog, 2018; Thelwall, 2018).

Each of these databases is built in a different way and has its unique selection criteria, indexing process, and therefore content. Web of Science is traditionally the most selective (Clarivate, 2020) and aims to concentrate on the highest impact academic journals. Scopus has broader coverage than Web of Science (Huang et al., 2020; Schotten et al., 2017), and Dimensions is the broadest of the three (Harzing, 2019; Visser, van Eck, & Waltman, 2021). Therefore, searching the same terms will produce a different result depending on which data source is searched.

## 3. Data sources and methods

### 3.1. Creating the datasets

We created four datasets of SDG 13-related research using the methods described by the distinct research groups as follows. In each case, we used all document types and limited records to the five-year time window 2015-19

- Elsevier – We used the Elsevier 2021 method which is the result of a two-step process. First, Scopus records were extracted using the search string defined as SDG 13 in the fourth update (Rivest et al., 2021). The resulting set of articles were then fed into an algorithm described by Rivest et al. (2021) that uses machine learning methods to enhance the original list.
- STRINGS – We used the search terms elicited through the methods described by Confraria et al. (2021) to query the titles, abstracts, and keywords of publications in Web of Science. The resulting publications were searched in approximately 4,000 clusters based on an article-level citation clustering described by Waltman & van Eck (2012). If a minimum 15% of any cluster contained our SDG13 related publications, then all publications in that cluster were included in our dataset. If the 15% threshold was not reached, then none of the publications in the cluster were included.

- SIRIS – We used the search strategy that combines keywords as described in the visual essay by SIRIS Academic (2020) to query the titles, abstracts, and keywords of publications in Web of Science.
- Dimensions – We used the SDG methods including the machine learning enhancements described by Wastl et al. (2020)

We ran the queries described above against versions of Web of Science and Dimensions housed in the database system of the Centre for Science and Technology Studies (CWTS) at Leiden University.

Our version of Web of Science includes the following five indexes: Science Citation Index – Expanded, Social Sciences Citation Index, Arts & Humanities Citation Index, and both editions of the Conference Proceedings Citation Index). Neither the Book Citation Index nor the Emerging Sources Citation Index were used because we do not have access to them.

Elsevier's International Centre for the Study of Research (ICSR) Lab kindly made the 2021 dataset available for the purpose of this study, which we used in combination with the CWTS version of Scopus.

### 3.2. Search term classification

We collected the search terms used by each of the methods and organised them into three groups using our own general knowledge of the field. We classified the search terms for each method according to the following criteria:

- General: Terms used in society, e.g., "temperature rise". The general public would use these terms.

- Policy: Terms that require knowledge of policy contents, e.g., "emissions trading". They do not include mere mention of policy, e.g., "Kyoto protocol" which would count as a general term.

- Technical: Terms that are technical in nature, so subject matter experts would use them. They either refer to a technology, e.g. "thermal energy storage", or require technical knowledge, e.g. "radiative forcing". Standard technologies, e.g. "solar panel", do not count – these would be considered general terms.

### 3.3. DOI analysis

First, we calculated the total number of publications for the five-year period 2015 to 2019 from each of the four datasets related to SDG 13. We then determined which of these records had a DOI. We subsequently used the DOI as the unique identifier when comparing the datasets. This means only records with DOIs were included in the comparisons.

### 3.4. Pairwise coverage comparisons

We performed pair-wise comparisons to examine the overlap between the four datasets. So, each dataset was compared with the other three, thereby making a total of six pairwise comparisons.

Because the four methods use different data sources, some of the surplus is due to differences in coverage between those data sources. For example, STRINGS uses Web of Science while Elsevier uses Scopus. Therefore, we sub-divided each surplus into two portions. One portion of the surplus was due to differences between the search strategies described by each of the methods, while the other portion was due to coverage differences between the data sources. We termed these 'surplus (method)' and 'surplus (coverage)' respectively.

### 3.5. Visualisation of the outputs

For each pairwise comparison, we then presented the results in the form of VOSviewer term maps. These maps visualise terms found in the titles and abstracts of the articles in two datasets and show the terms in different shades of colour depending on the frequency with which they occur in each dataset. Each term needed to appear a minimum of 70 times in publications retrieved from a pairwise combined set of search terms. The term maps offer an easy way to see which topics are over- or underrepresented in one dataset compared to the other.

## 4. Results

### 4.1. Publications with a DOI

Table 3 shows the total number of records identified by each of the four methods, and those that are associated with a DOI.

Table 3. SDG 13 records and DOIs for each selected method (2015-19)

| Method | SDG13 Publications | Publications with DOI | Share of publications with DOI |
|---|---|---|---|
| Elsevier 2021 | 214,369 | 195,734 | 91.3% |
| STRINGS | 166,528 | 156,010 | 83.7% |
| SIRIS | 177,154 | 164,800 | 93.0% |

| Dimensions | 205,190 | 203,447 | 99.2% |

The number of publications related to SDG 13 was relatively similar for each of the four methods chosen for the study. The largest set of records was found through the Elsevier 2021 method, although only about 5% larger than the Dimensions dataset. The SIRIS and especially STRINGS datasets were smaller, although the STRINGS dataset was still over three-quarters the size of the Elsevier 2021 total.

Each method had DOIs for at least 91% of its SDG 13 related publications. That meant we had a comparable set of publications associated with DOIs from the four methods for comparative study.

### 4.2. Comparison based on classification of search terms

The Elsevier 2021 method used an expansive list of keywords covering a range of topics related to climate change such as greenhouse gas emissions, and global warming. The list also extended into terms describing actions taken to address the problem such as policies and laws, but also addressed developments of resilient foods and agricultural methods. The term "legum* breed*" AND ("climate" or "drought" or "flood") is one of many technical terms related to food and agriculture in the context of climate action. These highly specific terms are designed to maximise recall while maintaining a high level of precision. There is also a considerable set of exclusion terms that use the AND NOT command, e.g., "Prehistoric Climate" and "blood". These are intended to exclude publications captured by the initial search terms, but that are not related to the current challenges surrounding climate action. The exclusions therefore improve precision of the dataset.

STRINGS used a lot of broad, simple terms, for example, "climate change". STRINGS but not SIRIS used the term "carbon economy", while SIRIS instead used more specific terms not employed by STRINGS, e.g., "carbon accounting", "carbon audit", "carbon credit", "carbon dividend", "carbon fee", and "carbon finance". STRINGS extracted search terms from lay documents including web forums and grey literature as well as policy documents and scientific publications to capture terms used by a broad section of society. The STRINGS surpluses due to method were all higher than those of other methods. We speculate that is the effect of the enhancement step that is based on the grouping of Web of Science into 4,000 clusters of publications related by citation links even in the absence of explicit use of keywords. The enhancement makes a decision about whether to include or exclude each of the Web of Science 4,000 topic clusters in the final dataset. If the cluster is selected, then all publications in that cluster are added. As the threshold for inclusion was set at 15%, it means that all publications in any cluster in which 15% of the records contain the seed keywords are included. However, the 15% inclusion threshold makes it possible that up to 85% of the records in a selected cluster did not in fact contain the keywords. This record 10.1080/09540091.2017.1279126 is exclusively retrieved by the STRINGS method, although it contains none of the STRINGS keywords. The explanation must be inclusion in a topic cluster selected because of the existence of other publications bearing the search terms. The citation-based grouping seems to have been more inclusive than the other methods in the study and emphasise recall over precision. Conversely, any topic cluster in which fewer than 15% records contain the seed keywords is excluded along with all its publications, even those that did contain the keywords. An example is 10.1371/journal.pone.0137275 which contains the term 'climate change' in its abstract. This term is included in the STRINGS seed keywords but the publication is not included in the final dataset. It must therefore have been excluded from STRINGS due to existence in a topic cluster mainly populated with less relevant papers. Therefore, even where the different methods used the same keywords, this enhancement step has produced discrepancies compared with the other methods.

Overall SIRIS used more than twice as many search terms as STRINGS, many of them technical. There were 54 'technical' search terms compared with only four in STRINGS. For example, "ocean acidification" and "radiative forcing" found thousands of records in SIRIS that did not appear in STRINGS. Sometimes SIRIS was restrictive, for example requiring the term "climate change" to be combined with others, i.e., "climate change" and ("policies" OR "education" OR "impact" OR "reduction" OR "warning" OR "planning" OR "strategy" OR "mitigation"). Conversely, the simple mention of "greenhouse gas" qualified publications for inclusion in SIRIS, while STRINGS required the same term to be combined with another term like "emission", "reduction", or "changing climate". The technical terms used by SIRIS contributed to large numbers of publications in the SIRIS surpluses against all the other methods.

Dimensions used only 45 search terms, most of them general. However, these were searched against a larger database. The Dimensions method also employed the proximity search in almost all the search terms so that phrases that included certain words in close proximity would be found. For instance, 'Climate related hazards'~3 will also find articles that contain 'hazards related to climate change' in their titles or abstracts. The advantage is that publications that include phrases used in the context of climate action could be returned rather than only finding an exact phrase.

The number of search terms used by each method is shown by type in Table 4. The Elsevier 2021 method used mainly general and technical terms plus about 14% policy related terms.

The STRINGS method used a high proportion of general terms, but the remainder were almost all policy related with very few technical terms. The SIRIS method was far more specific with about a quarter of the search terms policy related and a quarter technical in nature.

Table 4. Search term classification

| Method | General | Policy | Technical | Total |
|---|---|---|---|---|
| Elsevier 2021 | 210 (46%) | 62 (14%) | 186 (41%) | 458 |
| STRINGS | 70 (71%) | 24 (24%) | 4 (4%) | 98 |
| SIRIS | 119 (52%) | 55 (24%) | 54 (24%) | 228 |
| Dimensions | 34 (76%) | 9 (20%) | 2 (4%) | 45 |

The full list of terms along with their classification is available in Zenodo.

### 4.3. Comparison based on overlap of publications

In each pairwise comparison the set of overlapping records is shown in the central portion of the Venn diagram. Only publications with a DOI are used in order to make these comparisons. The records found in one dataset but not the other can be termed surplus. In the sample diagram (Figure 1), the two portions to the left are each included in dataset A, but not in dataset B and therefore comprise the dataset A surplus. As the methods use different bibliographic databases (Table 2), the surplus can be sub-divided into the portion of the surplus due to the differences in method, and the portion due to differences in coverage.

The reader may consider these comparisons as a Venn diagram flattened into a stacked horizontal bar as shown in Figure 1.

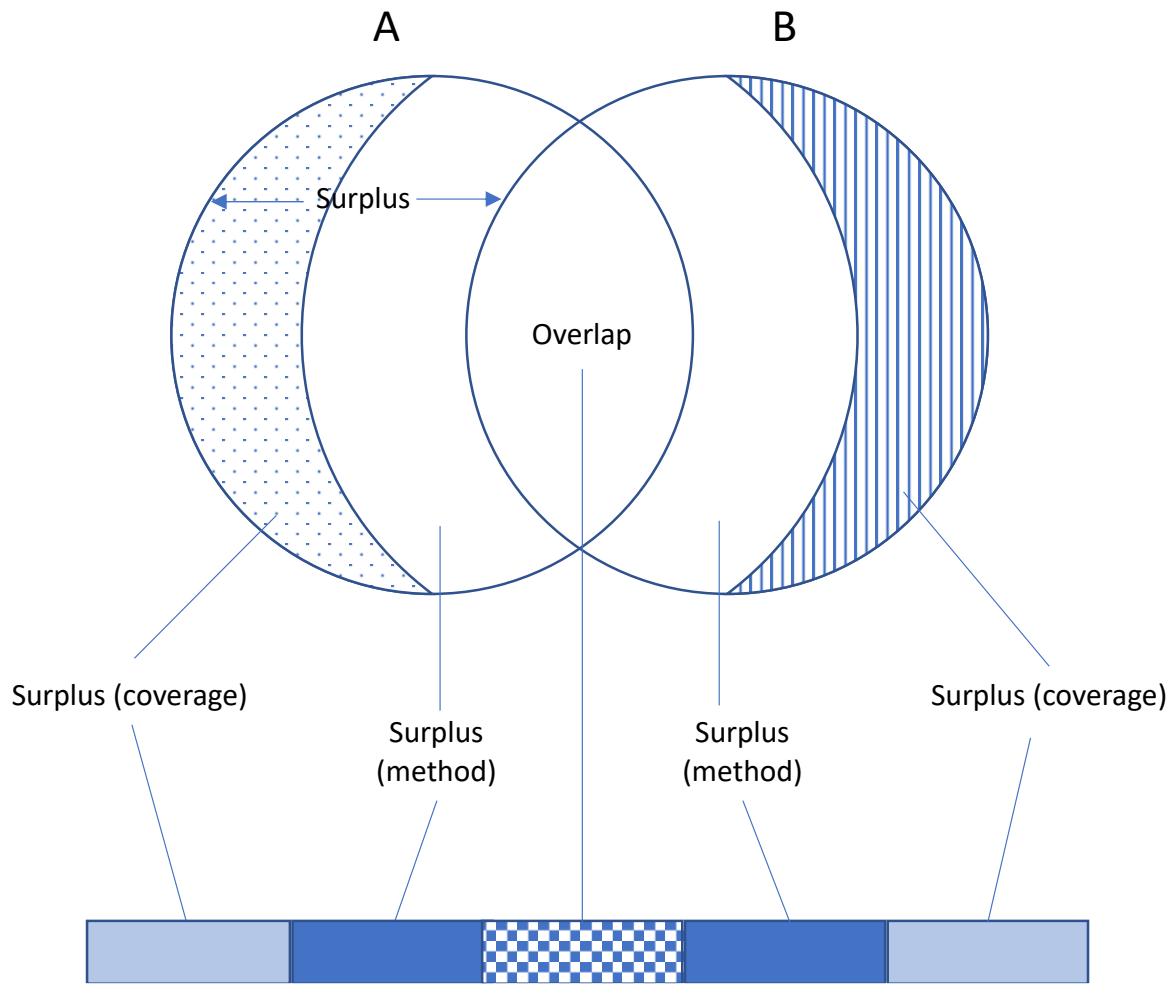

Figure 1. Key to overlap and surplus

Figure 2 shows pair-wise coverage comparisons of SDG 13 related publications between the four different methods. Each bar is labelled with the two datasets compared. The number of records in each portion of the pairwise comparison is shown in Table 5.

The first comparison shows that 56,043 publications were found by both Elsevier 2021 and STRINGS, and these are represented by the central portion of the bar. Immediately to the left of the overlap are 79,613 records found in the Elsevier 2021 dataset, but not STRINGS because of the difference between the two SDG 13 search strategies (surplus due to method). The far-left portion of the bar represents a further 36,544 records in the Elsevier 2021 dataset, but not in the STRINGS set because these records are not found in Web of Science (surplus due to coverage).

Likewise, the other end of the bar shows 2,949 STRINGS publications that were not found through the Elsevier 2021 method because they are not indexed in Scopus. The remaining 97,018 STRINGS publications were not found in the Elsevier 2021 dataset due to differences between the two search strategies.

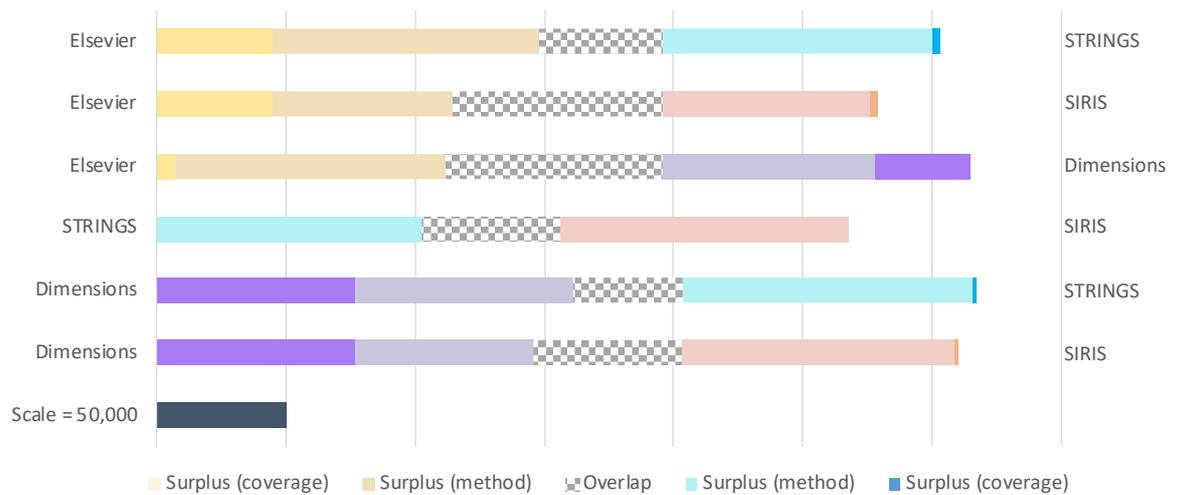

Figure 2. Number of overlapping and surplus publications between methods

The largest overlap (29.2%) was between the Elsevier 2021 and SIRIS methods (Table 5), while the lowest (13.4%) was between Dimensions and STRINGS. Overlap means that both methods in the comparison retrieved the same publications. The range of agreement is surprisingly low, indeed no two methods compared show a high degree of overlap.

Surplus due to method ranged from 22.2% to 41.6%. These are publications that were found by one method but not the other, where the discrepancy was attributed to the method of identifying the publications. The high level of surplus due to method demonstrates the large disagreements between all four methods.

The surplus due to database coverage was very low (maximum 1%) for both methods that used Web of Science (STRINGS and SIRIS), confirming the selective coverage of Web of Science. Conversely, Dimensions showed in one case (vs. SIRIS) that almost a quarter (24.6%) of the combined records in the pair were in its surplus due coverage. These are publications found by one method but not the other where the discrepancy is attributable to the coverage of the data source. There was no surplus due to coverage for the STRINGS vs SIRIS comparison because both used the Web of Science database.

Table 5. Number and share of overlapping and surplus publications

| Method A | Surplus (coverage) | Surplus (method) | Overlap | Surplus (method) | Surplus (coverage) | Method B |
|---|---|---|---|---|---|---|
| Elsevier | 44,764 | 102,702 | 48,269 | 104,792 | 2,949 | STRINGS |
|  | 14.8% | 33.8% | 15.9% | 34.5% | 1.0% |  |
| Elsevier | 44,764 | 69,502 | 81,469 | 80,421 | 2,910 | SIRIS |
|  | 16.0% | 24.9% | 29.2% | 28.8% | 1.0% |  |
| Elsevier | 7,103 | 104,613 | 84,019 | 82,587 | 36,831 | Dimensions |
|  | 2.3% | 33.2% | 26.7% | 26.2% | 11.7% |  |
| STRINGS | 0 | 102,564 | 53,446 | 111,354 | 0 | SIRIS |
|  | 0.0% | 38.4% | 20.0% | 41.6% | 0.0% |  |
| Dimensions | 76,389 | 84,629 | 42,429 | 112,522 | 1,059 | STRINGS |
|  | 24.1% | 26.7% | 13.4% | 35.5% | 0.3% |  |

| Dimensions | 76,389 | 68,933 | 58,125 | 105,494 | 1,181 | SIRIS |
|---|---|---|---|---|---|---|
| | 24.6% | 22.2% | 18.7% | 34.0% | 0.4% | |

Sample DOIs for each group are available via a link in Zenodo.

### 4.4. Comparison based on topical focus of publications

Analysis of the resulting publications visualised through VOSviewer showed terms extracted from the titles and abstracts of publications and grouped by co-occurrences in publications. Each comparison shows two maps. The first map allows us to assign broad descriptive phrases like 'energy problem' to clusters of papers with the most relevant and frequently occurring terms represented by different colours. In Figure 3, the first map groups terms into three distinct, colour coded fields related to energy problem (green), CO2 problem (blue), and climate change (red).

The second map shows for each term which of the two methods of identifying SDG 13 related research captured more publications that use the term. Each bubble represents a term, and the colour of the bubble reflects the score of the term. Terms that occurred more frequently in the publications identified by the first method have a negative score and appear over a blue bubble, while terms that occurred more frequently in publications identified by the second method have a positive score and appear over red bubbles. Terms that appear over the faded colour bubbles occurred evenly in publications identified by both methods. In Figure 5, the second map shows us that the terms related to the CO2 problem tend to occur more frequently in the Elsevier dataset. Meanwhile, the terms related to the energy problem appeared more frequently in the Dimensions dataset.

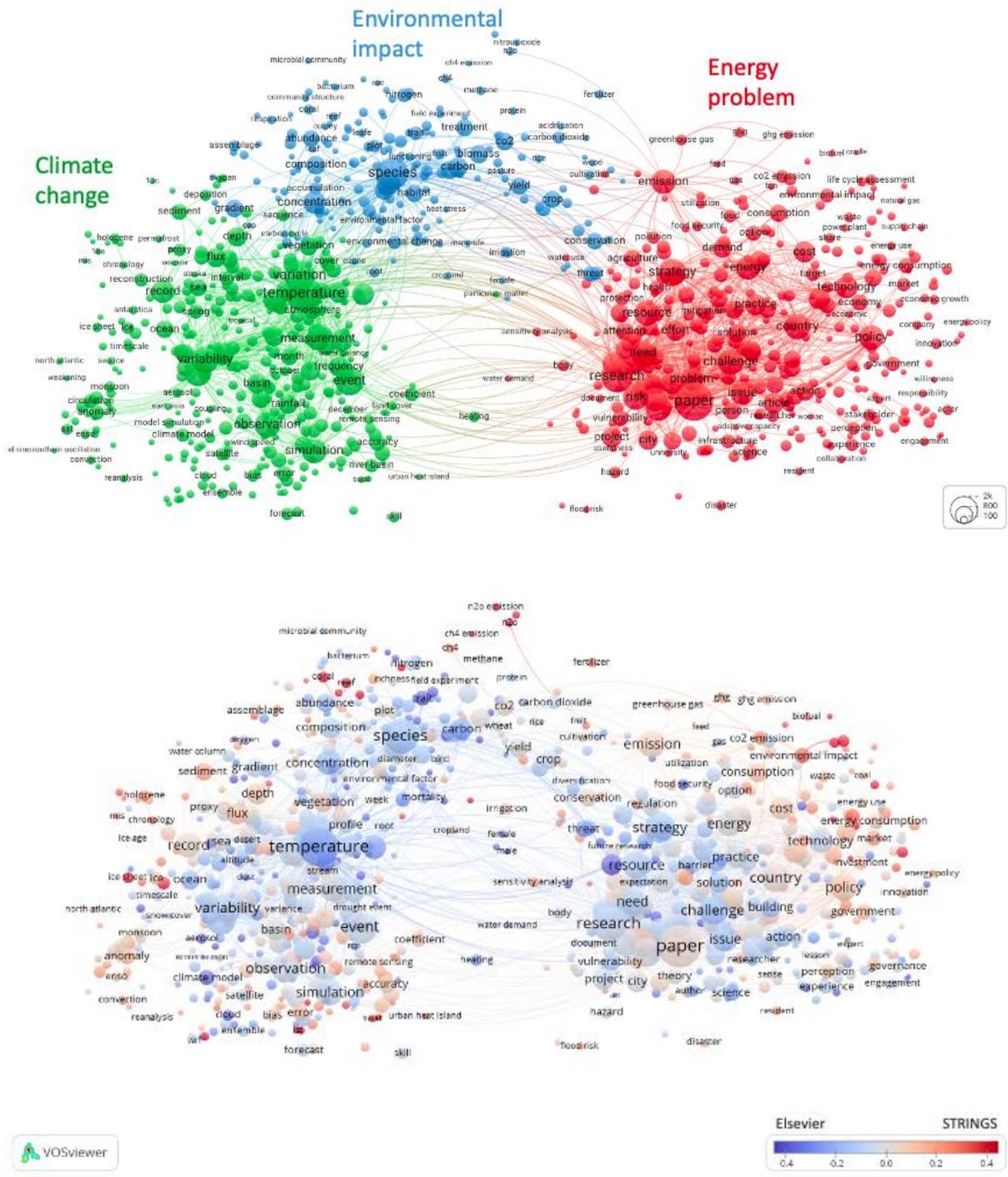

Figure 3. Elsevier vs STRINGS - Click to navigate interactive live map

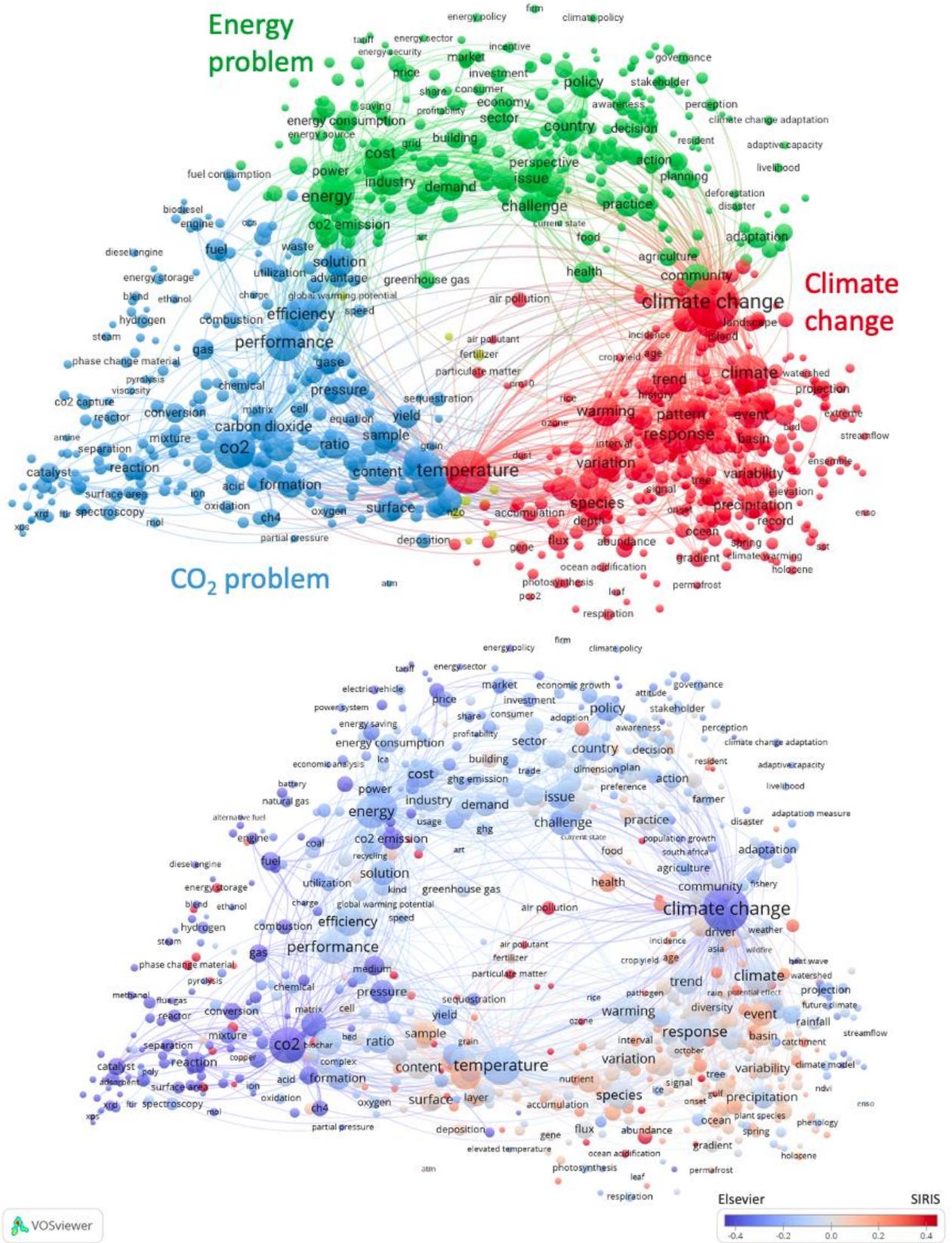

Figure 4. Elsevier vs SIRIS Click to navigate interactive live map

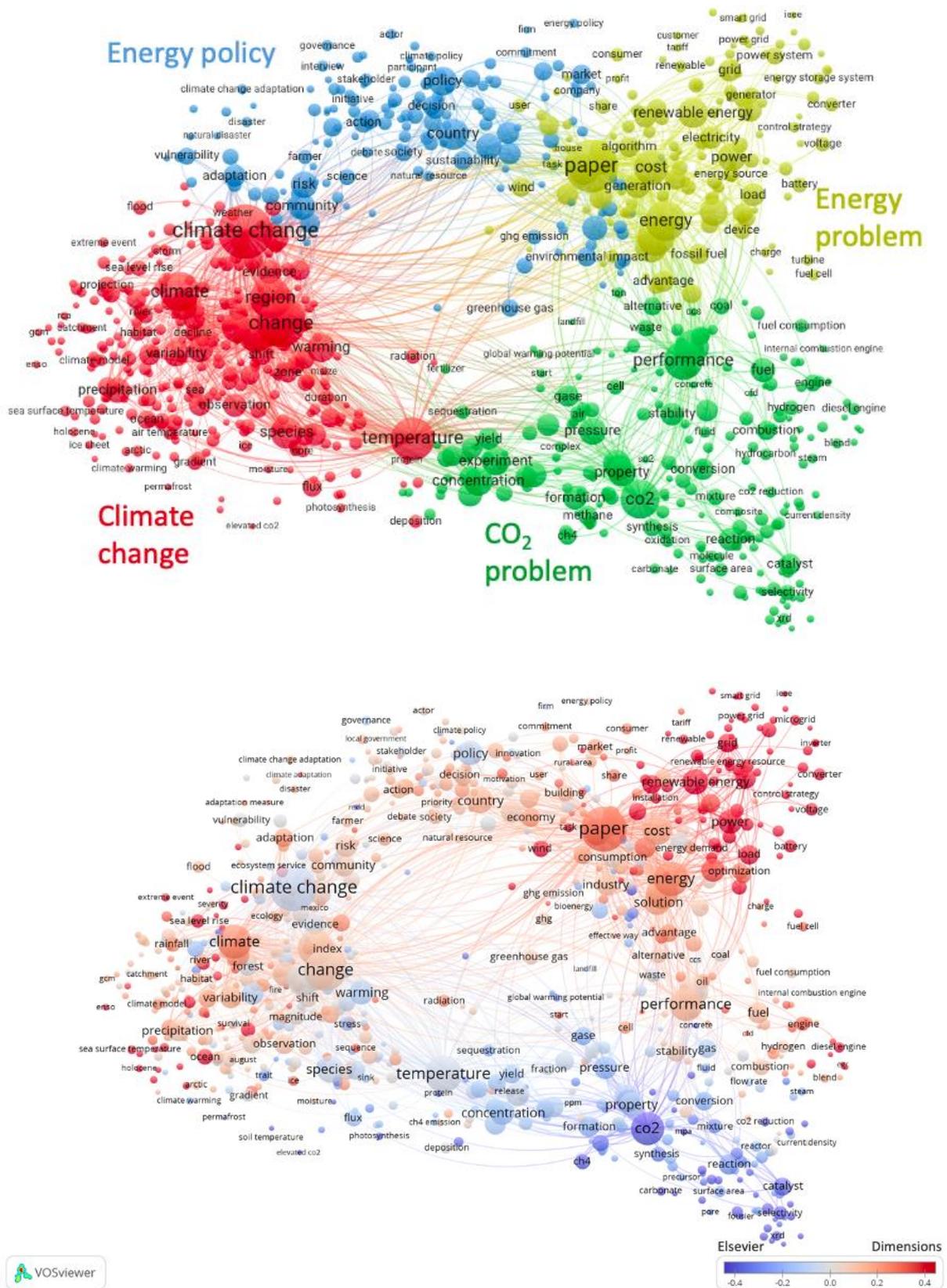

Figure 5. Elsevier vs Dimensions Click to navigate interactive live map

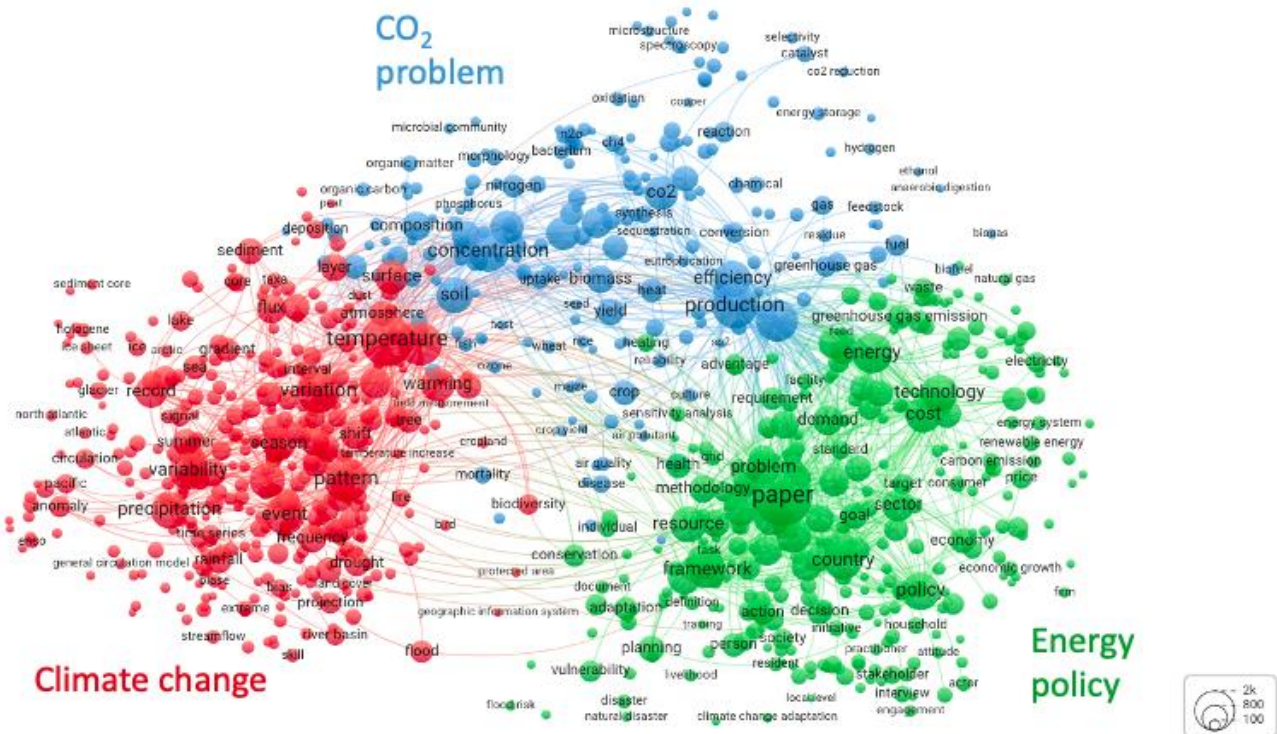

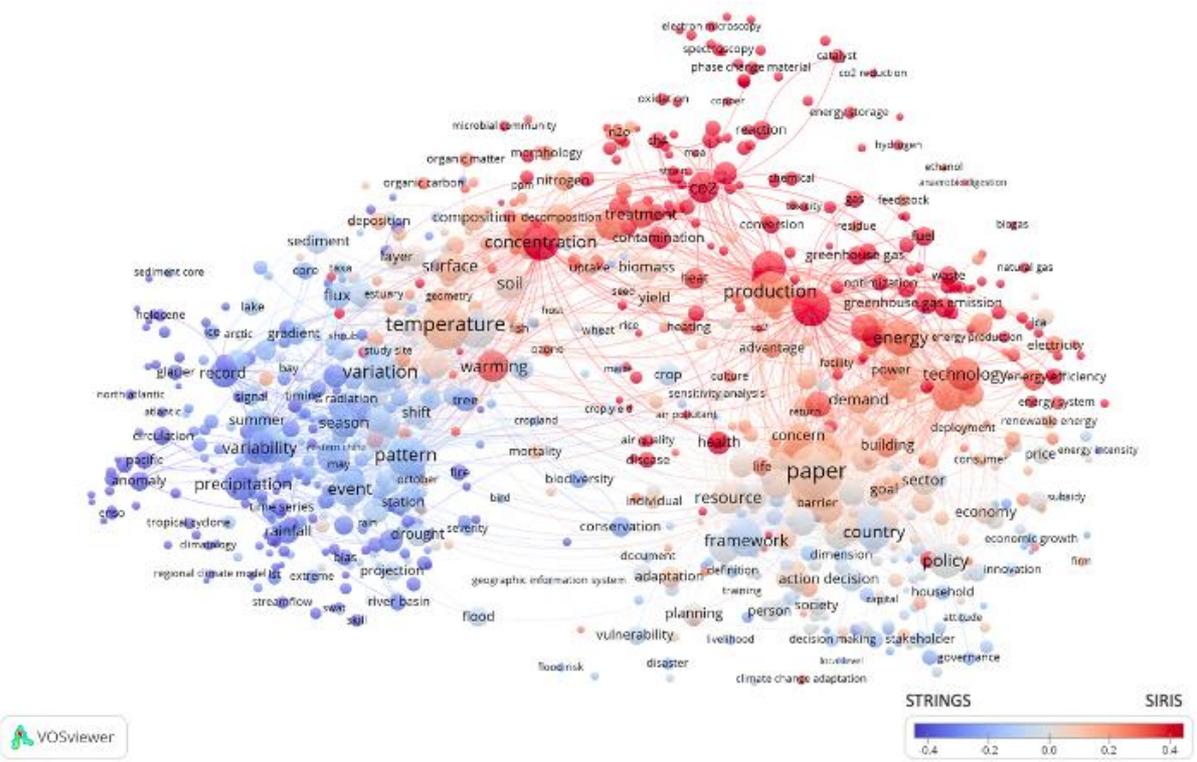

Figure 6. STRINGS vs SIRIS - Click to navigate interactive live map

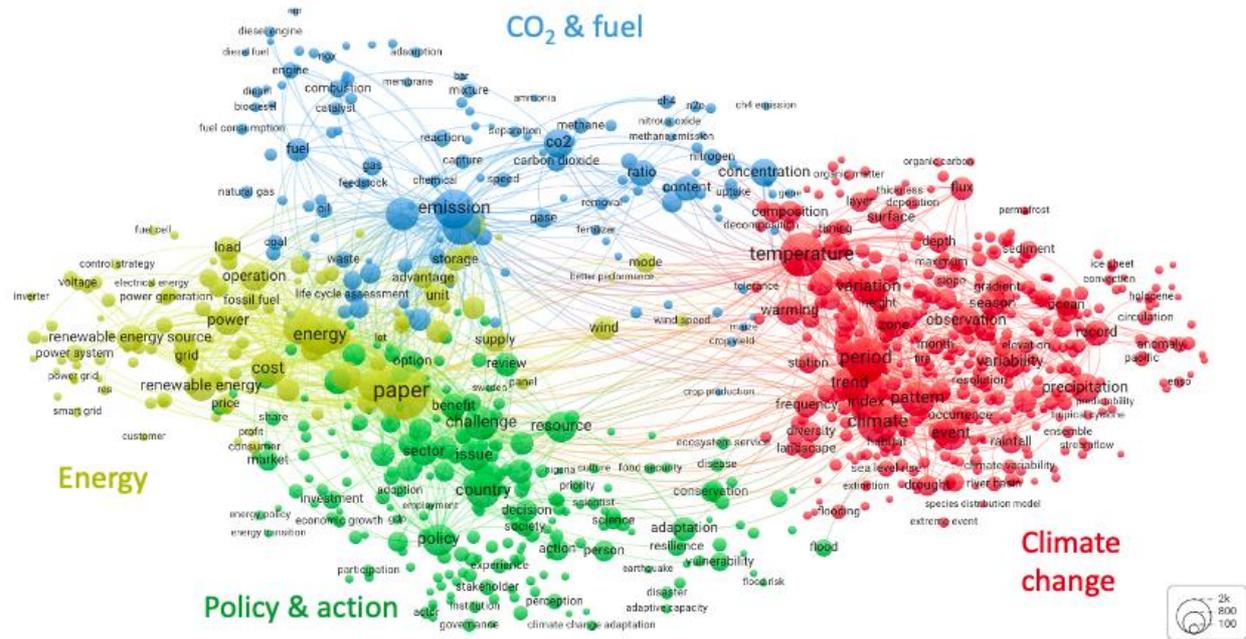

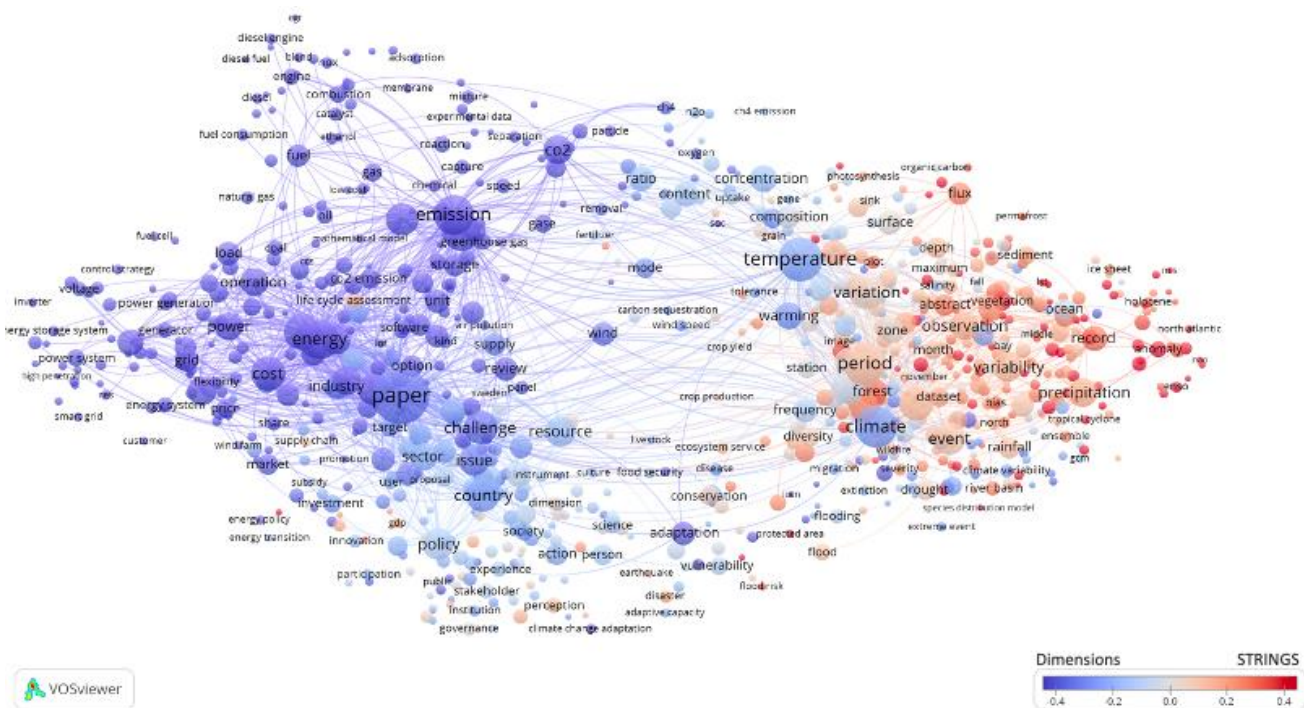

Figure 7. Dimensions vs STRINGS - Click to navigate interactive live map

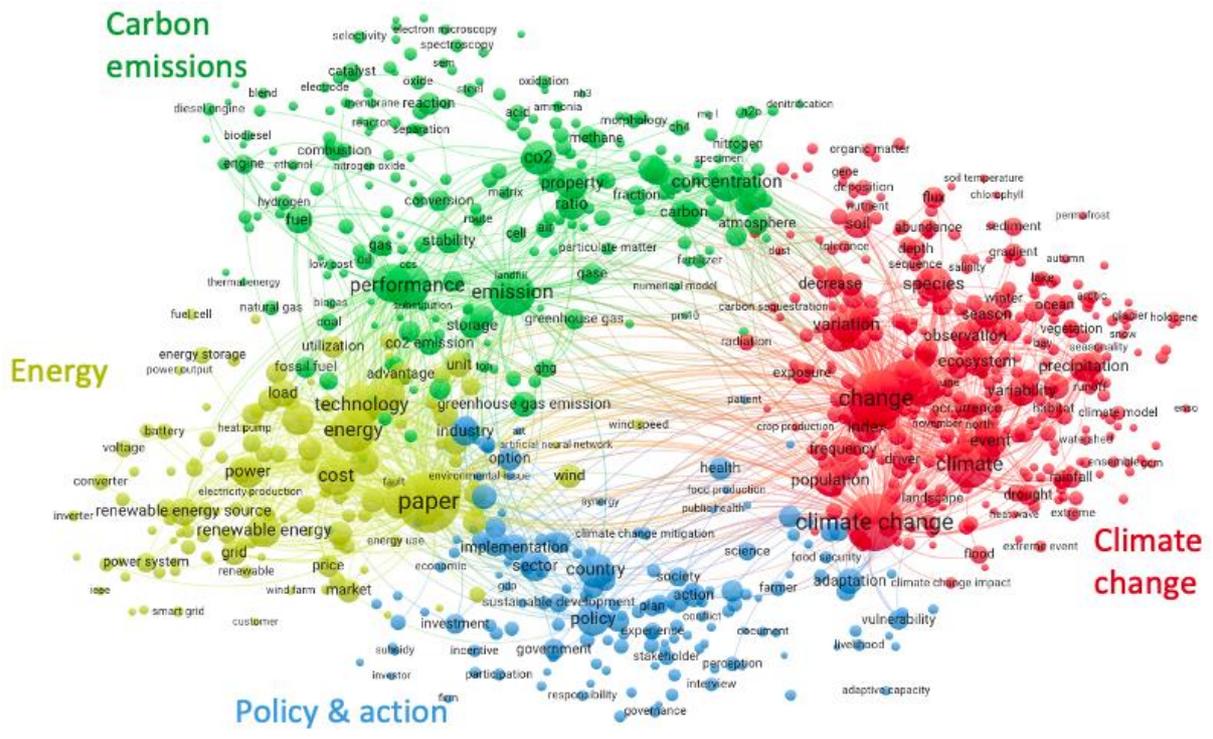
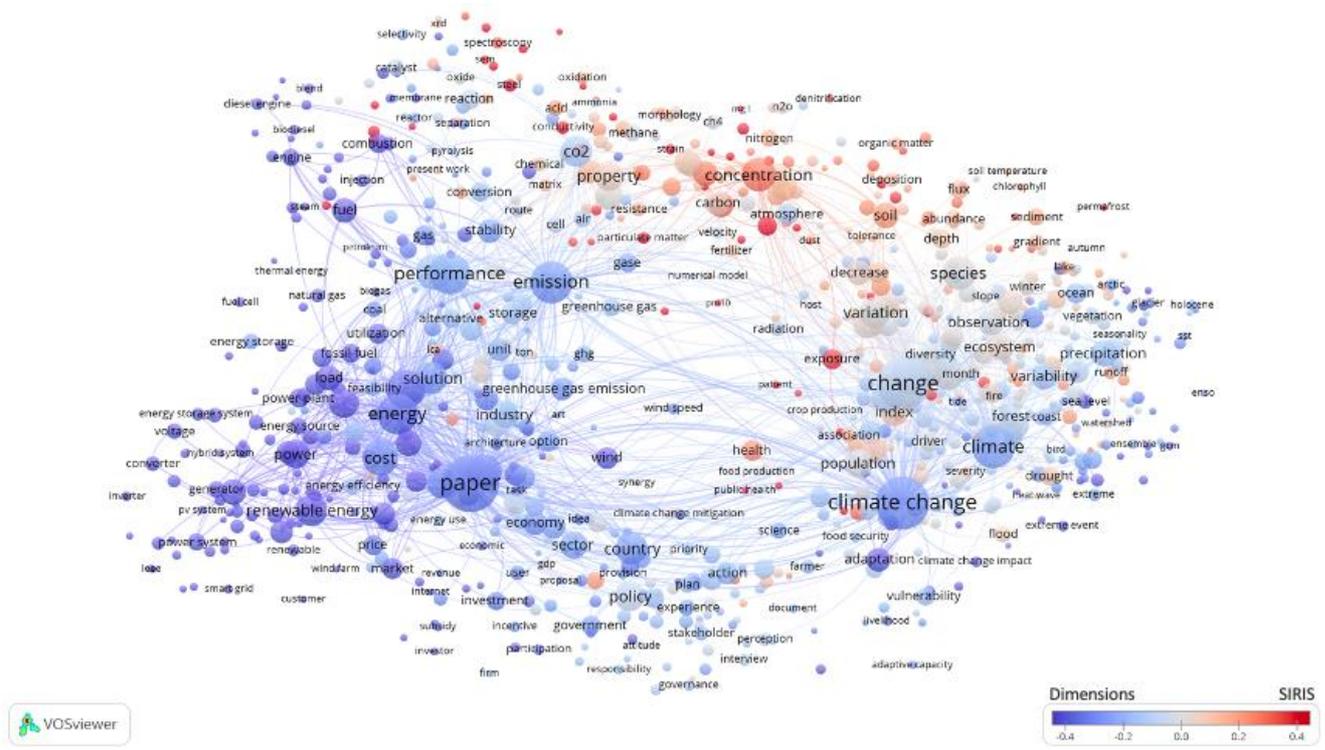
Figure 8. Dimensions vs SIRIS - Click to navigate interactive map

The four methods studied have all identified SDG 13-related publications but each with a discernible topical emphasis. Each map presents the terms most frequently found in a combined set of publications for the methods compared. The terms may appear in publications identified by both methods, but the colour indicates which method identified them more frequently.

The first three VOSviewer maps show terms expressed in research publications identified by the Elsevier method compared with the other three methods. In these maps, there is no clearly discernible pattern. In the comparison with STRINGS, the Elsevier method has perhaps identified papers that more frequently use terms related to the environment and climate. This might be due to the inclusion of large numbers of related search terms in the Elsevier seed keywords. Alternatively, Elsevier's machine learning enhancement might have trained the search engine to identify these publications, or perhaps Scopus has indexed more papers in this field than the other databases.

In the three comparisons with STRINGS, patterns are easier to see. The colouring of terms related to climate change indicated they occurred more frequently in the publications identified by STRINGS than the other methods, although this distinction was less clear in the comparison with Elsevier. STRINGS used broader, more encompassing search terms, e.g. 'climate change', than the other methods. This broad search strategy might have contributed to recall of a larger set of publications that contained related terms. STRINGS then introduced more publications to its dataset by adding all papers in clusters related by citation links. We did not quantify these additions, but entire clusters of publications were added if 15% or more records in the cluster contained the keywords. We assume this approach added many publications on climate change and contributed to their prominence in the maps. As STRINGS uses Web of Science as its data source, it is also possible that database indexes publications on climate change more frequently than Scopus or Dimensions. If that were the case, it would at least partially explain the prominence of the records in the STRINGS comparisons. However, SIRIS also used Web of Science and in the pairwise comparison, STRINGS clearly found climate change publications more frequently.

The SIRIS method appears to have retrieved publications more focused on the technical nature of carbon emissions. SIRIS used a relatively large number of keywords, and they were highly technical in their nature. SIRIS avoided broad terms like 'carbon emissions' but instead used 32 more specific terms containing the word 'carbon' such as 'orbiting carbon observatory' and 'personal carbon trading'. Construction of SIRIS search terms was supported by natural language processing and we speculate that has resulted in the more frequent inclusion of publications with technical terms as seen in the maps. Again, database coverage would provide an alternative explanation if Web of Science indexed more technical publications than the other databases. However, the comparison with STRINGS, which also used Web of Science, showed SIRIS to find more technical publications, making database coverage a less likely explanation.

Dimensions demonstrated some prominence in publications with terms related to energy and policy. Dimensions searches the term 'renewable energy' which retrieves a large quantity of publications, whereas the other methods require that term to be used in combination. An explanation might partially lie in the interpretation of the subject matter experts of the term 'climate action'. Experts might differ in their emphasis with some focusing on the 'action' part while others may see the term more synonymous with climate change in general. If the experts used in the Dimensions method wanted to focus on the action, then it would make sense to choose terms containing verbs such as 'reduce emissions' and 'limit global temperature rise'. This point of view might also lead experts to select the names of agreements and forums among its search terms as places where action is discussed. Elsevier took a conscious decision not to include 'renewable energy' as a stand-alone search term for SDG 13 publications to reduce overlap with SDG 7 (Affordable and clean energy). Dimensions is the broadest of the three data sources used and its larger journal coverage might also have contributed to its large surpluses against the other methods.

## 5. Discussion

In this study we compared the publication sets retrieved by four different methods of identifying research related to SDG 13: Climate action. Each method begins by selecting relevant keywords from the SDG goal and its related targets. These keywords are then combined to create a query that is searched on a bibliographic database. Each method then enhanced its results in different ways. The resulting set of publications from the four methods overlapped very little, given that they all started with the same task.

Overlap is defined as publications that were retrieved by two methods directly compared with each other. Any publications found by one method but not the other are discrepancies. The fact that each method comprises multiple stages means that we cannot easily determine the source of any discrepancy. The method in effect becomes a black box. Our inputs are the keywords and the resulting publication set the output. Discrepancies between the publication sets may be the result of any stage of the methods compared. Those designing methods of identifying SDG related research should be encouraged to open the black box by publishing each element of their method so end users can choose from a more informed perspective, see Table 6.

Table 6 – Elements of the black box

| Element | Description |
| --- | --- |
| Seed keyword selection | Description of source documents and how keywords were selected |
| Use of experts | Type of expertise, time invested, and instructions given |
| Operationalisation of search strategies | List of concatenated search terms |
| Reference sets used to assess recall | Description of reference sets and how they were constructed |
| Random sampling of the reference sets | Sample publications from the reference sets |
| Source database selection | Database, edition, and any additional parameters used |
| Enhancement techniques | Detailed description of methods used to enhance the dataset |
| Random sampling of retrieved publications | Samples of publications retrieved before and after enhancement |

To begin with, the set of seed keywords selected by the four methods were very different with up to a ten-fold difference in the number and type of keywords used. This level of difference is of primary interest and raises questions around interpretation of the goal. Each method used experts or analysts with familiarity of the topic to select the keywords so why were they so different? The manual element of building the search strings is crucial because human decision controls which terms are included and how they are combined. People with deep knowledge of the field will be likely to produce terms of a more technical nature. These terms will increase recall while minimising less relevant publications found by broader terms. However, none of the methods provide the identity of the experts, say how they were selected, how much time they spent, the precise instructions they were given, or how they resolved differences in expert opinion.

This missing information is key because experts might differ in their precise field of expertise. Some will be more knowledgeable about technical details of the problems surrounding SDGs and select more technical terms. Others might be more familiar with the details of the climate agreements and choose more policy related terms. Even experts with similar levels of knowledge will have their own views as to what is relevant and what is not. For instance, is research on nuclear energy relevant or not in the context of SDG 13? What about medicine's role in mitigating the effects of climate change on health? Each expert will have their own views on these and other questions, and the choices they make is a likely source of divergence in publication retrieval.

Similarly, the combination of keywords is of great importance and construction of the search queries varied between the four methods. Each method used a combination of broad, collective search terms that increase recall, and highly technical terms designed to maximise precision. Broad terms are good for recall but raise the prospect of contaminating the final dataset with less relevant papers. Elsevier 2021 and SIRIS used many more terms than the other methods. They included highly specific, technical terms that found publications in more concentrated fields. Search strategies that use many narrow, specific terms might produce precise datasets, but require many more such terms to build up a corpus of publications.

The creation of thematic datasets undoubtedly involves an element of subjectivity due to the human dimension. For topics as complex as the SDGs, this is even more challenging due to the diverse nature of subject matter experts. Their expertise will always be different making it difficult for them to reach reliable consensus on what research is relevant and what is not. Under such circumstances, operationalising a specific definition or interpretation of SDG 13 in the form of a reference dataset is critical. This reference set of publications can be used to test the recall, i.e., what share of the reference set is retrieved by the implemented search queries? The query can be systematically expanded until a certain minimum threshold of recall is reached. The query should be tweaked during this process to keep precision above a defined cut-off point. Such reference sets can be made up of specialist journals, specialised research groups, or publication clusters highly relevant to the target literature. Both the selection of the reference set and the recall rate will influence the outcome and overlap with other methods. Even if difference methods started with the same interpretation of an SDG, they would still produce different results because of discrepancies between their operationalisation processes. Unfortunately, we know too little about how each method operationalised their searches limiting our ability to compare them. This requires further investigation to reveal the causes of the low level of overlap between the methods reported in this study as well as to guide future work on the development of such datasets.

The four methods used three different databases between them. Web of Science is the most selective of the three and aims to capture scholarly literature from high impact sources. Meanwhile Scopus has become more inclusive, and Dimensions searches an even larger corpus of literature. That STRINGS and SIRIS had relatively small coverage surpluses against the other methods confirmed expectations as Web of Science indexes fewer publications than the other databases. Conversely, Dimensions found tens of thousands of additional publications because they are not indexed in Scopus or Web of Science. This is expected because we know that Dimensions covers many publications not indexed in Scopus or Web of Science (Visser et al., 2021).

The databases have different coverage policies and the publications indexed therefore vary. So even running the same search query on different databases will produce different datasets depending on the emphasis of coverage. Consequently, differences in topic emphasis identified in this study might easily be due to the choice of data source rather than nuances of the search queries. It should be noted that the SIRIS approach was designed to be database agnostic. We applied the SIRIS search strategy to the Web of Science but would expect different results if the same strategy were applied to other bibliographic databases. One method of isolating the impact of database is to run the keyword searches of one method across different databases. The Bergen Group (Armitage et al., 2020a) made an attempt at this by translating Elsevier's 2019 keyword search strings into Web of Science syntax. However, this requires great skill, is not always possible, and might raise questions over differences in understanding between the original author and the translator – a common problem in language translation (van Nes, Abma, Jonsson, & Deeg, 2010). In our study, we isolated the impact of data source by separating the surplus into that caused by data source and the remaining portion that we could attribute the rest of the method.

Finally, all the methods enhanced their datasets but each in a different way. Elsevier 2021 and Dimensions used experts in multiple rounds of relevance checking and then used machine learning algorithms to increase recall. STRINGS added or removed publications depending on whether they were in a relevant topic cluster of publications. SIRIS employed natural language processing at an early stage to produce a long and specific list of technical keywords. The effect of these enhancements could be assessed in a series of controlled studies that only assess the effect of the enhancement. For example, Elsevier has documented (Rivest et al., 2021) a comparison between its pre- and post-enhancement datasets and some details behind the machine learning algorithms. We could potentially use the enhancement technique of one of the methods and apply it to the keywords of each method in the comparison. At this stage, we do not have access to all the details of all the enhancement methods and therefore did not attempt analysis of the enhancements or their impacts.

We have established that great differences between datasets exist but what do they mean? The differences described above will necessarily compound one another to produce the datasets and it is perhaps not surprising that they overlap so little. Publications found by one method but not another might be intentional. Each of the methods involves human decision based on interpretation of the intended outcome, selection of relevant keywords, and construction of the search strings. There may be legitimate differences between the understanding and aims of one group of experts and another. To properly identify the source of the differences between datasets, we need to analyse each stage of the identifications methods in isolation to better understand their contribution to the overall differences. In the present study, the comparison between STRINGS and SIRIS eliminates the effect of the database because they both used Web of Science.

As the study of identifying SDG research intensifies, the methods used will come under greater scrutiny. Any groups designing such methods should therefore fully and publicly document their approach step by step. It is important for readers to know details of the search strategy such as which keywords were used, who selected them, and how they were combined into search strings. Database selection is also important because it will determine which records are available for retrieval and impact the size of the final dataset. Any enhancements should be described in full, and algorithms deposited in a public repository. The more details provided, the easier the method will be to justify. This is an area of growing interest and peers will be pleased to help improve on methods.

During the course of this study, many questions were raised that could be the subject of follow up studies. The four methods compared were complex and to fully understand their differences would require a systematic controlled comparison at each step. Other limitations to the study should also be considered.

This study focused entirely on one SDG and any conclusions drawn can only be interpreted in that context. As we did not gain a good understanding of the reasons for the discrepancies between methods, we cannot predict whether they would be similar if we used a different SDG in the case study. Broader studies could look at multiple SDGs to detect any patterns.

We chose the DOI as the unique identifier to compare overlapping coverage between data sources because of the extent of its use in academic publishing. Most records in the SDG datasets have a DOI. However, a small fraction of records was not included in the comparisons because they did not have a DOI.

There is also a small share of publications with discrepancy in the publication year between different bibliographic databases. Both Scopus and Web of Science assign the publication year of an article as the official date of publication of the journal issue. Dimensions assigns the publication year based on the date the article was first available – usually the online version (Digital Science, 2021). Consequently, our datasets may exclude a small number of publications from Elsevier, STRINGS, and SIRIS from the latter part of the time window while including the same records in Dimensions. Likewise, some records at the beginning of the time window may be included in Elsevier, STRINGS, and SIRIS, but excluded from Dimensions. However, the overall effect of discrepancies in publication year is likely to be small.

# 6. Conclusions

Each of the four methods compared has attempted to identify research related to climate action and produced largely different results. Their search strategies were created using human judgement and ranged from broad and simple to technical and focused. Between the four methods, three different bibliographic databases were used, each with their own unique coverage. Finally, in some cases, machine learning and other artificial intelligence techniques were applied to enrich the final publication datasets.

These findings support those by earlier work by Armitage et al. (2020a) and build further by comparing four methods and visualising their outputs in the context of their search strategies. This study also shows the relative contribution of search strategy and data source to the different publication datasets produced.

Using broader data sources to apply the search strategies increases the number of documents returned simply because of the larger coverage. Dimensions comprises more documents than either Scopus or Web of Science and might offer benefits to some methods, especially those aiming to find relevant literature beyond the constraints of highly selective journal literature. The STRINGS method makes a deliberate attempt to find search terms from grey literature and web forums. It might therefore be logical to apply these search terms against the broadest possible data source, i.e., Dimensions.

The search strategy, use of subject matter experts, and data source vary between the four methods. Each method therefore produces a different set of publications related to SDG 13. The fact that we have several different answers to the same questions produces a major implication. The overlap in publications found by these different methods is too low to be adopted by policy makers without careful method selection. The choice of method will potentially define the resulting dataset more than any other factor. Any comparison between research entities should use the same method of identifying publications. As more studies on research into climate change appear in the literature, readers should avoid the temptation to draw hasty conclusions. Published assessments of SDG-related research should state the method used along with other variables such as the time period and data source. The method used is an important influencer of the number and type of resulting publications.

# Acknowledgements

Ton van Raan and Ludo Waltman for invaluable input and expert guidance throughout the study, Ed Noyons for help with creating the STRINGS and SIRIS datasets, Elsevier ICSR Lab for providing the Elsevier (2021) dataset related to SDG 13. Stephanie Faulkner (Elsevier), Ed Noyons and Ismael Rafols (STRINGS), Francesco Massucci (SIRIS), Juergen Wastl (Dimensions), and two anonymous reviewers for extensive feedback and helpful comments on earlier versions of the article.

## Competing interests

## Funding

No funding was sought or received for this project.

## Data availability

The search terms presented in table 4 and subsets of the DOIs from table 5 are available in Zenodo (Purnell, 2022).

## References

Adams, J. (1998). Benchmarking international research. *Nature*, *396*(6712), 615–618. https://doi.org/10.1038/25219

Archambault, E., Campbell, D., Gingras, Y., & Lariviere, V. (2009). Comparing of Science Bibliometric Statistics Obtained From the Web and Scopus. *Journal of the American Society for Information Science and Technology*, *60*(7), 1320–1326. https://doi.org/10.1002/asi.21062

Armitage, C., Lorenz, M., & Mikki, S. (2020a). Mapping scholarly publications related to the Sustainable Development Goals: Do independent bibliometric approaches get the same results? *Quantitative Science Studies*, *1*(3), 1092–1108. https://doi.org/10.1162/qss_a_00071

Armitage, C., Lorenz, M., & Mikki, S. (2020b). *Replication data for: Mapping scholarly publications related to the Sustainable Development Goals: Do independent bibliometric approaches get the same results?* (V1 ed.; U. of Bergen, Ed.). V1 ed. https://doi.org/doi:10.18710/98CMDR

Association of DutchUniversities. (2019). SDG-Dashboard: Impact of the Dutch Universities. Retrieved June 21, 2021, from https://vsnu.nl/en_GB/sdg-dashboard-english.html

Baas, J., Schotten, M., Plume, A., Côté, G., & Karimi, R. (2020). Scopus as a curated, high-quality bibliometric data source for academic research in quantitative science studies. *Quantitative Science Studies*, *1*(1), 377–386. https://doi.org/10.1162/qss_a_00019

Blasco, N., Brusca, I., & Labrador, M. (2021). Drivers for Universities' Contribution to the Sustainable Development Goals: An Analysis of Spanish Public Universities. *Sustainability*, *13*(1). https://doi.org/10.3390/su13010089

Clarivate. (2020). Web of Science Journal Evaluation Process and Selection Criteria - Web of Science Group. Retrieved September 5, 2020, from https://clarivate.com/webofsciencegroup/journal-evaluation-process-and-selection-criteria/

Confraria, H., Noyons, E., & Ciarli, T. (2021). *Countries research priorities in relation to the Sustainable Development Goals*.

Digital Science. (2021). Which publication dates does Dimensions use? Retrieved from https://dimensions.freshdesk.com/support/solutions/articles/23000019982-which-publication-dates-does-dimensions-use-

Duran-Silva, N., Fuster, E., Massucci, F. A., & Quinquillà, A. (2019). *A controlled vocabulary*

*defining the semantic perimeter of Sustainable Development Goals*. https://doi.org/10.5281/ZENODO.4118028

Elsevier SciDev.Net. (2015). *Sustainability science in a global landscape*. Retrieved from https://www.elsevier.com/__data/assets/pdf_file/0018/119061/SustainabilityScienceReport-Web.pdf

European Commission. (2020). CORDIS - EU Research Results. Retrieved March 23, 2021, from About CORDIS website: https://cordis.europa.eu/about/en

Garfield, E. (1972). Citation analysis as a tool in journal evaluation. *Science*, *178*(4060), 471–479. https://doi.org/10.1126/science.178.4060.471

Gläser, J., Glänzel, W., & Scharnhorst, A. (2017). Same data—different results? Towards a comparative approach to the identification of thematic structures in science. *Scientometrics*, *111*(2), 981–998. https://doi.org/10.1007/s11192-017-2296-z

Harzing, A.-W. (2019). Two new kids on the block: How do Crossref and Dimensions compare with Google Scholar, Microsoft Academic, Scopus and the Web of Science? *Scientometrics*, *120*(1), 341–349. https://doi.org/10.1007/s11192-019-03114-y

Herzog, C., Hook, D., & Konkiel, S. (2020). Dimensions: Bringing down barriers between scientometricians and data. *Quantitative Science Studies*, *1*(1), 387–395. https://doi.org/10.1162/qss_a_00020

Hirsch, J. E. (2005). An index to quantify an individual{\textquoteright}s scientific research output. *Proceedings of the National Academy of Sciences*, *102*(46), 16569–16572. https://doi.org/10.1073/pnas.0507655102

Hook, D. W., Porter, S. J., & Herzog, C. (2018). Dimensions: Building Context for Search and

Evaluation. *Frontiers in Research Metrics and Analytics*, *3*, 23. https://doi.org/10.3389/frma.2018.00023

Huang, C.-K., Neylon, C., Brookes-Kenworthy, C., Hosking, R., Montgomery, L., Wilson, K., & Ozaygen, A. (2020). Comparison of bibliographic data sources: Implications for the robustness of university rankings. *Quantitative Science Studies*, *1*(2), 445–478. https://doi.org/10.1162/qss_a_00031

International Science Council. (2015). *Review of Targets for the Sustainable Development Goals: The Science Perspective (2015)*. Retrieved from https://council.science/publications/review-of-targets-for-the-sustainable-development-goals-the-science-perspective-2015

Jayabalasingham, B., Boverhof, R., Agnew, K., & Klein, L. (2019). Identifying research supporting the United Nations Sustainable Development Goals. *Mendeley Data*, *1*. https://doi.org/10.17632/87txkw7khs.1

Jetten, T. H., Veldhuizen, L. J. L., Siebert, M., Ommen Kloeke, A. E. E. van, & Darroch, P. I. (2019). *An explorative study on a university's outreach in the field of UN Sustainable Development Goal 2*. (June), 0–22. https://doi.org/10.18174/476199

Körfgen, A., Förster, K., Glatz, I., Maier, S., Becsi, B., Meyer, A., … Stötter, J. (2018). It's a Hit! Mapping Austrian Research Contributions to the Sustainable Development Goals. *Sustainability*, *10*(9). https://doi.org/10.3390/su10093295

Nakamura, M., Pendlebury, D., Schnell, J., & Szomszor, M. (2019). *Navigating the Structure of Research on Sustainable Development Goals*. Retrieved from https://clarivate.com/webofsciencegroup/campaigns/sustainable-development-goals/

Nature. (2021). Tracking 20 leading cities' Sustainable Development Goals research. *Nature*. https://doi.org/10.1038/d41586-021-02406-9

Provençal, S., Campbell, D., & Khayat, P. (2021). *Provision and analysis of key indicators in research and innovation. Policy brief J, Research trends on the Sustainable Development Goals (SDGs) and alignment with SDG 17 through international co-publications : focusing on SDGs 12–15 plus 6 (Planet)*. https://doi.org/doi/10.2777/03227

Purnell, P. J. (2022). *A comparison of different methods of identifying publications related to the United Nations Sustainable Development Goals: Case Study of SDG 13: Climate Action*. Retrieved from https://zenodo.org/record/6861335

Rafols, I., Noyons, E., Confraria, H., & Ciarli, T. (2021). Visualising plural mappings of science for Sustainable Development Goals (SDGs). *SocArxiv*. https://doi.org/10.31235/osf.io/yfqbd

Rivest, M., Kashnitsky, Y., Bédard-Vallée, A., Campbell, D., Khayat, P., Labrosse, I., … James, C. (2021). *Improving the Scopus and Aurora queries to identify research that supports the United Nations Sustainable Development Goals (SDGs) 2021 Version 4*. https://doi.org/10.17632/9sxdykm8s4.4

Schotten, M., Aisati, M. El, Meester, W. J. N., Steiginga, S., & Ross, C. A. (2017). A brief history of Scopus: The world's largest abstract and citation database of scientific literature. *Research Analytics: Boosting University Productivity and Competitiveness through Scientometrics*. https://doi.org/10.1201/9781315155890

SIRISAcademic. (2020). Is EU-funded research and innovation evolving towards topics related to the Sustainable Development Goals? Retrieved March 23, 2021, from Visual Essay website: http://science4sdgs.sirisacademic.com

Thelwall, M. (2018). Dimensions: A competitor to Scopus and the Web of Science? *Journal of Informetrics*, *12*(2), 430–435. https://doi.org/10.1016/j.joi.2018.03.006

TimesHigherEducation. (2021a). Impact Rankings 2021: methodology. Retrieved June 14, 2021, from https://www.timeshighereducation.com/world-university-rankings/impact-rankings-2021-methodology

TimesHigherEducation. (2021b). Impact rankings 2021. Retrieved July 17, 2021, from https://www.timeshighereducation.com/rankings/impact/2021/overall#!/page/0/length/25/sort_by/rank/sort_order/asc/cols/undefined

United Nations. (2014). *The Road to Dignity by 2030: Ending Poverty, Transforming All Lives and Protecting the Planet, Synthesis Report of the Secretary-General on the Post-2015 Agenda*. Retrieved from http://www.un.org/ga/search/view_doc.asp?symbol=A/69/700

United Nations. (2017). Global indicator framework for the Sustainable Development Goals and targets of the 2030 Agenda for Sustainable Development. Retrieved March 17, 2021, from https://unstats.un.org/sdgs/indicators/Global Indicator Framework after 2021 refinement_Eng.pdf

van Nes, F., Abma, T., Jonsson, H., & Deeg, D. (2010). Language differences in qualitative research: is meaning lost in translation? *European Journal of Ageing*, *7*(4), 313–316. https://doi.org/10.1007/s10433-010-0168-y

van Raan, A. (1999). Advanced bibliometric methods for the evaluation of universities. *Scientometrics*, *45*(3), 417–423. https://doi.org/10.1007/BF02457601

Vanderfeesten, M., & Otten, R. (2017). *Societal Relevant Impact : Potential analysis for Aurora-Network university leaders to strengthen collaboration on societal challenges*.

https://doi.org/10.5281/ZENODO.1045839

Vanderfeesten, M., Spielberg, E., & Gunes, Y. (2020). *Survey data of "Mapping Research Output to the Sustainable Development Goals (SDGs)"* https://doi.org/10.5281/ZENODO.3813230

Visser, M., van Eck, N., & Waltman, L. (2021). Large-scale comparison of bibliographic data sources: Scopus, Web of Science, Dimensions, Crossref, and Microsoft Academic. *Quantitative Science Studies*, *2*(1), 20–41. https://doi.org/10.1162/qss_a_00112

Waltman, L., & van Eck, N. J. (2012). A new methodology for constructing a publication-level classification system of science. *Journal of the American Society for Information Science and Technology*, *63*(12), 2378–2392. https://doi.org/10.1002/asi.22748

Waltman, L., van Eck, N. J., van Leeuwen, T., Visser, M. S., & van Raan, A. F. J. (2011). Towards a new crown indicator: an empirical analysis. *Scientometrics*, *87*(3), 467–481. https://doi.org/10.1007/s11192-011-0354-5

Wastl, J., Hook, D. W., Fane, B., Draux, H., & Porter, S. J. (2020). *Contextualizing Sustainable Development Research*. https://doi.org/10.6084/m9.figshare.12200081